\documentclass[seceq]{ptptex}

\usepackage{wrapft}

\usepackage{latexsym}
\usepackage{amsmath}
\usepackage{wasysym}
\usepackage{amsfonts}

\def\pe{\prime}
\def\3s{{s \choose 3}}
\def\4s{{s \choose 4}}
\def\5s{{s \choose 5}}
\def\6s{{s \choose 6}}

\def\12{\frac{1}{2}}
\def\fr{\frac}
\def\ft{\footnote}

\def\pr{\partial}

\def\prd{\partial \cdot}

\def\be{\begin{equation}}
\def\ee{\end{equation}}

\def\a{\alpha} 
\def\b{\beta}  
\def\g{\gamma} 

\def\d{\delta} 

\def\e{\epsilon}

\def\h{\eta}

\def\l{\lambda}
\def\L{\Lambda}
\def\m{\mu}
\def\n{\nu}

\def\vf{\varphi}

\def\pb{{\bar\psi}}

\def\dsll{\not {\! \pr}}
\def\psisl{\not {\! \! \psi}}

\def\ssl{\not {\! \cal S}}

\def\ssl{\not {\! \cal S}}

\def\cA{{\cal A}}
\def\cB{{\cal B}}

\def\cD{{\cal D}}
\def\cE{{\cal E}}
\def\cF{{\cal F}}

\def\cL{{\cal L}}

\def\cR{{\cal R}}
\def\cS{{\cal S}}

\def\cW{{\cal W}}

\title{%        %You can use \\ for explicit line-break.
Low-spin models for higher-spin Lagrangians
}
\subtitle{}    %Use this when you want a subtitle.

\author{%       %Use \scshape  for the family name.
Dario \textsc{Francia}\footnote{e-mail: francia@fzu.cz} 
}

\inst{%         %Affiliation, neglected when [addenda] or [errata].
Institute of Physics - Academy of Sciences of the Czech Republic \\
Na Slovance 2  CZ-18221 Prague -  Czech Republic\footnote{Address after September 30, 2011:
Centro Studi e Ricerche E. Fermi,
Piazza del Viminale 1, I-00184 Roma - Italy, and
Scuola Normale Superiore and INFN, Piazza dei Cavalieri 7, I-56126 Pisa - Italy.}
}

\abst{%         %This abstract is neglected when [addenda] or [errata].
Higher-spin theories are most commonly modelled on the example of spin 2. While this is appropriate
for the description of free irreducible spin-s particles, alternative options could be equally interesting. 
In particular Maxwell's equations provide the effective model for maximally reducible 
theories of higher spins inspired by the tensionless limit of the open string. For both options, as well as 
for their fermionic counterparts, one can extend the analogy beyond the equations for the gauge potentials, 
formulating the corresponding  Lagrangians in terms of higher-spin  curvatures. The associated non-localities 
are effectively  due to the elimination of auxiliary fields and do not modify the spectrum. 
Massive deformations of these theories are also possible, and in particular in this contribution we  
propose a generalisation of the Proca Lagrangian for the Maxwell-inspired geometric theories.}

\begin{document}
\maketitle

\section{Introduction}

In a couple of previous works\cite{dariocrete, dariotripl} we proposed to interpret
higher-spin\ft{See \cite{HSP} for reviews and  recent progress in various directions.}
Lagrangians involving metric-like curvature tensors\cite{dwf} as the result of integrating away auxiliary
fields from unconstrained local actions\cite{fs3, fms1}. Our main goal was to provide a rationale for the appearance of  
non-local operators in the corresponding theories, while at the same time suggesting a set of effective rules 
for their manipulation.  

Previous explorations of the possible role of curvatures in the dynamics of higher-spins in metric form\cite{fs1, fs2, fms1} 
were mostly devoted to  generalising  the linearised Einstein-Hilbert (or Fierz-Pauli) equations for spin $2$, along with their 
massive deformations\cite{dario07, darioValencia}.
However, the Lagrangian found in Ref.~\citen{dariotripl}, being for any spin $s$ the square of the corresponding 
higher-spin curvature, should be more properly regarded as a generalisation of Maxwell's Lagrangian.
The physical difference between the results of Refs.~\citen{fs1, fs2, fms1} and the equations written in Ref.~\citen{dariotripl}
is that while the former describe the free theory of a single particle, the latter propagate a number of irreducible
representations of different spins, in a sense that will be recalled in the following.
It thus seems that the two basic examples of low-spin gauge fields also provide two different models
for possible  higher-spin theories: the Einstein-Hilbert linearised equations being a model for 
the  propagation of a single particle of arbitrary spin, while more direct generalisations of Maxwell's theory 
describe in a compact form whole sets of propagating particles. The corresponding roles for fermions are played
by the Rarita-Schwinger theory for irreducible representations and by the Dirac equation for reducible ones.

Our general motivation is related to the possibility that non-linear deformations of higher-spin 
curvatures\ft{See for instance Ref.~\citen{manvelyan}.}, to be properly included within a Lagrangian
framework, might improve our insight into interactions, while the apparent non-locality of the free theory,
as already stressed, would represent only a spurious effect due to the elimination
of auxiliary fields and would not interfere with the particle interpretation of the weakly coupled theory.
In this respect, since \emph{any} unconstrained local Lagrangians can be reinterpreted in terms of curvatures
along the lines of Refs.~\citen{dariocrete, dariotripl} here reviewed, it is possible to select in principle several
different free theories, some of which might allow a simpler generalisation to the non-linear level than others, 
which motivates our interest in exploring alternative forms. 

Certainly, any reducible theory can be ultimately diagonalised to better display its irreducible content, 
leading to vertices involving single-particle fields. What is not at all obvious in principle is that the form and the geometric
interpretation of the couplings themselves should appear more natural when written in terms of the irreducible fields. 
Moreover, while here we concentrate on two extremal cases, in a sense, namely maximally reducible and  maximally 
irreducible theories, other options describing an \emph{intermediate} particle content are also possible
in principle. In particular the class of geometric solutions found in Ref.~\citen{fms1}, including  its simplest 
representatives first constructed in Refs.~\citen{fs1, fs2}, should be reconsidered, in our opinion, in this wider perspective.
We plan to investigate  these issues in future works.

We discuss bosons and fermions  in Section \ref{sec:bosons} and 
Section \ref{sec:fermions} respectively, 
while Section \ref{sec:massive} contains an exposition of the massive theory including the form
of the generalised Proca Lagrangians that we propose here. In Section $5$ we indicate some directions 
for future work.

%%%%%%%%%%%%%%%%%%%%%%%%%%%%%%%%%%%%%%%%%%%%%%%%%%%%%%%%%%%%%%%%%%%%%

\section{Massless bosons}\label{sec:bosons}

%%%%%%%%%%%%%%%%%%%%%%%%%%%%%%%%%%%%%%%%%%%%%%%%%%%%%%%%%%%%%%%%%%%%%

%%%%%%%%%%%%%%%%%%%%%%%%%%%%%%%
\subsection{Irreducible case}\label{sec:irrbosons}
%%%%%%%%%%%%%%%%%%%%%%%%%%%%%%%

 As a prototype for equations propagating a single, irreducible representation of the Poincar\'e group we take
the Fierz-Pauli equation for the linearised graviton, \cite{FP} in its two equivalent forms:
\be \label{FP}
\begin{split}
& \Box \, h_{\, \m \n} \, - \, \pr_{\, \m} \, (\prd h)_{\, \n} \, - \, \pr_{\, \n} \, (\prd h)_{\, \m} \, + \, \pr_{\, \m} \, \pr_{\, \n} \, h^{\, \a}{}_{\, \a} \, = \, 0\, , \\
& \h^{\, \a \b } \, \cR_{\, \a \b; \, \m \n} \, = \,  0 \, ,
\end{split}
\ee 
where $\cR_{\, \a \b; \, \m \n}$ is the linearised Riemann tensor, written in the convention of Ref.~\citen{dwf} with explicit
symmetry between indices in a given group. 
The most successful generalisation of the first equation in  \eqref{FP}   to the case of spin $s > 2$ is the Fronsdal 
equation:\ft{Here $\vf$ is a rank-$s$ symmetric tensor whose indices are omitted for simplicity.
Products of different tensors are symmetrised with the minimal number of terms, with no weight factors. 
Lorentz traces are denoted by ``primes'' or by numbers in square brackets,  while ``$\prd$'' stands for a divergence. 
A list of combinatorial rules needed to exploit the benefits of this notation can be found in 
Refs.~\citen{fs1, fms1, dario07}.  We use the ``mostly-plus'' space-time metric in $d$ dimensions.}
 \cite{}
\be \label{F}
\cF \, \equiv \, \Box \, \vf \,  - \, \pr \, \prd \vf \, + \, \pr^{\, 2} \, \vf^{\, \pe} \, = \, 0 \, ,
\ee
describing the propagation of a massless particle of spin $s$, under the assumption that the  
linearised diffeomorphisms of $h_{\, \m \n} $,  $\d h_{\, \m \n} =  \pr_{\, \m}  \L_{\n} +  \pr_{\, \n} \L_{\m}$,
be generalised to abelian gauge transformations with a traceless parameter: $\d \vf =  \pr \L$, $\L^{\, \pe} \equiv  0$.
\cite{fronsdal} 
The main  limitation of \eqref{F} is that it does not admit a direct interpretation in terms of higher-spin
curvatures, thus apparently implying that the equivalence of the two forms in \eqref{FP}
only represents a fortunate accidental property of lower spins. However, proper generalisations
of the geometric form of \eqref{FP} actually exist, originally proposed in Ref. \citen{fs1, fs2} and 
subsequently elaborated upon in Refs. \citen{fms1}, \citen{dario07}, \citen{dariocrete, dariotripl}. 
To better understand the systematics and the meaning of the corresponding construction it is convenient to first recall  
the minimal unconstrained extension of \eqref{F}. This involves an auxiliary field $\a$ transforming proportionally
to the trace of the gauge parameter: $\d \a \, = \, \L^{\, \pe}$, allowing to generalise \eqref{F}  to the equation\cite{fs2, st}
\be \label{A}
\cA \, \equiv \, \cF \, - \, 3 \, \pr^{\, 3} \, \a \, = \, 0 \, ,
\ee
clearly describing the same dynamics. The associated Lagrangian\cite{fs3} involves an additional auxiliary field $\b$
of rank $s-3$, acting as a Lagrange multiplier for the double-trace of $\vf$. The latter in fact needs to be absent 
in the Fronsdal formulation, in order to allow the derivation of \eqref{F} from an action principle. 
Although manifest from the construction, it is important to stress that neither $\a$ nor $\b$ mix with the physically 
propagating polarisations (nor do they carry additional degrees of freedom); it is thus clear that integrating
over these fields we get \emph{the same} theory. The corresponding ``effective'' Lagrangian does not involve any constraints, 
and because of that admits an interpretation in terms of higher-spin curvatures, although possibly leading to
equations more involved than
the simple condition of vanishing Ricci tensor \eqref{FP}. In Ref.~\citen{dariocrete} it was conjectured, and there verified for spin $3$ and 
spin $4$, that the resulting generalised Ricci tensor for spin $s$ should be identified with
\be \label{RicciS}
\cA_{\, \vf} \, = \, \cF \, -\, 3 \, \pr^{\, 3} \, \a_{\, \vf} \, = \, 
\sum_{k=0}^{n+1} \, a_k \fr{\pr^{\, 2k}}{\Box^{\, k}}\,
\cF_{n+1}^{\, [k]} \, ,
\ee
where
\be
a_k \, = \, (-1)^{k+1} \, (2\, k\,  -\, 1)\, 
\{ \fr{n\, + \, 2}{n\, - \, 1} \, \prod_{j=-1}^{k-1}\, \fr{n \, + \, j}{n \, - \, j \, + \, 1} \} \, ,
\ee
and where the tensors $\cF_{n+1}$ effectively compute successive traces of the curvatures 
$\cR_{\m_1 \cdots \m_s; \, \n_1 \cdots \n_s} \equiv \cR$ 
through the relations
\be \label{kinetic}
   \cF_{n+1} \, = 
     \begin{cases}
    \fr{1}{\Box^{\, n}} \, \cR^{\, [n+1]} \, & \, s \, = \, 2\, (n \, + \, 1) \, , \\
    \fr{1}{\Box^{\, n}} \, \prd \cR^{\, [n]} \, & \, s \, = \, 2\, n \, + \, 1 \, ,
    \end{cases}
\ee
also displaying how \eqref{RicciS} actually reduces to the standard local forms for $s=1, 2$.
The full form of the kinetic tensor \eqref{RicciS} was first obtained in Ref.~\citen{fms1} requiring 
that the corresponding Lagrangian  define the correct propagator, 
equivalent to that of the associated local unconstrained theory. The point of view
advocated in Ref.~\citen{dariocrete} provides an independent, \emph{a priori} justification of \eqref{RicciS} 
(although more complete checks are still to be performed), also clarifying the meaning of the associated non-localities, originally justified in order to
balance the higher derivatives appearing in the definition of higher-spin curvatures.\ft{The same procedure applied to 
alternative unconstrained theories such as those in Ref.\citen{buch} is expected to give the same result, although
through a different series of steps related to the different structure of auxiliary fields involved.}
Let us also mention that, alternatively to the non-local Lagrangians, the condition of vanishing trace of the curvature was also proposed as 
a consistent equation (although higher-derivative and non-Lagrangian) for irreducible massless spin $s$.\cite{bbl} 

To summarise, the description of free irreducible massless particles of arbitrary spin
is modelled on the spin-$2$ example at different levels. In  Fronsdal's theory
the Fierz-Pauli equation, viewed as an equation for the gauge potential,
is kept identical in form at the price of assuming algebraic constraints and of losing contact with the geometric form 
of the same equation. After removing the constraints by means of auxiliary fields
one obtains a theory where the gauge potential $\vf$ displays the same properties needed in
the construction of higher-spin curvatures in their metric form. Since the additional fields introduced
do not mix with the propagating polarisations one can consistently integrate over them and the
resulting effective kinetic tensor  \eqref{RicciS} can be written in terms of curvatures, in a form that manifestly 
generalises the geometric version  of the Fierz-Pauli equation \eqref{FP}.

As an additional comment, let us stress that the effective compensator $\a_{\, \vf}$ defined in \eqref{RicciS}
is not simply characterised by its gauge transformation, since this latter requirement would admit several 
different solutions\cite{fms1}. Rather, $\a_{\, \vf}$ is uniquely determined asking that $\cA_{\, \vf}$
possesses the same properties as its local ancestor \eqref{A}, i.e. that it satisfies $\cA_{\, \vf}^{\, \pe \pe} \equiv 0$
and $\prd \cA_{\, \vf} \, - \, \12 \, \pr \, \cA_{\, \vf}^{\, \pe} \equiv 0$.

%%%%%%%%%%%%%%%%%%%%%%%%%%%%%%%
\subsection{Reducible case}\label{sec:irrbosons}
%%%%%%%%%%%%%%%%%%%%%%%%%%%%%%%

We propose to consider the two forms of Maxwell's theory
\be \label{M}
\begin{split}
& \Box \, A_{\, \m} \, - \, \pr_{\, \m} \, \prd A \, = \, 0\, , \\
& \pr^{\, \n} \, F_{\, \n \m} \, = \,  0 \, ,
\end{split}
\ee
where $F_{\, \n \m}$ is the usual field-strength, as a model for describing the free propagation of \emph{reducible} higher-spin fields. 
By ``reducible'' we mean, somehow loosely speaking, that the resulting equations of motion actually propagate
a number of particles of different spins, but no ghosts.

For higher-spins, the proper generalisation of the first of \eqref{M} is the ``reduced'' triplet equation\cite{triplets, fs2, st, ftrev}
\be \label{triplet}
{\cal{T}} \, \equiv \, \Box \, \vf \, - \, \pr \, \prd  \vf \,+ \, 2 \, \pr^{\, 2} \, D = \, 0\, ,
\ee
where $\vf$ is the spin-$s$ potential while $D$ is an auxiliary field subject to $\d D = \prd \L$.
The origin of \eqref{triplet} is especially interesting, being the outcome of taking the 
tensionless limit in the free Lagrangian of open string field theory, after the elimination 
of an additional field $C$ also emerging in the same limit. (Which justifies our naming for \eqref{triplet}.) 
Eq. \eqref{triplet} describes the propagation of one massless particles for each of the spins
$s, \, s-2, \, s-4,  \, \cdots $, representing the \emph{maximally reducible} particle content of $\vf$ 
(as opposed to \eqref{F} or \eqref{A}, propagating only the single irrep of maximal spin), 
compatible with the absence of ghosts. 

Once again, while \eqref{triplet} nicely generalises the first of \eqref{M} it is not clear at first glance
how to formulate a possible analogue of the second of \eqref{M}. 
Observing that, due to the presence of the auxiliary field $D$, eq. \eqref{triplet} looks closer
in spirit to \eqref{A} rather than \eqref{F}, by analogy with our previous discussion 
one can look for the effective theory for \eqref{triplet} obtained after the integration over $D$.\cite{dariotripl} 
The outcome, as expected, is a non-local version of \eqref{triplet}
whose translation in terms of curvatures is indeed remarkably simple (especially if compared to
\eqref{RicciS}) and looks
\be \label{Tphi}
{\cal{T}_{\, \vf}} \, = \, \Box \, \vf \, - \, \pr \, \prd  \vf \, + \, 2 \, \pr^{\, 2} \, D_{\, \vf} \, = \, \fr{1}{\Box^{s - 1}} \, \prd^{\, s} \, \cR \, = \, 0 \, ,
\ee 
where the single divergence of the field-strength in \eqref{M} gets replaced by $s$ divergences of the corresponding
curvature for spin $s$, with the appropriate inverse power of the D'Alembertian operator.\ft{By 
comparison with the result of Ref.\citen{bbl} we conjecture that there might exist a local analogue
of \eqref{Tphi} in the form $\prd \cR \, = \, 0$, propagating the same degrees of freedom as \eqref{triplet} and 
no ghosts, despite being a higher derivative equation.} The analogy
between the spin-$1$ model and its string-inspired higher-spin avatar can be further appreciated
observing that, restoring indices for better clarity, the Lagrangian giving rise to \eqref{Tphi},
\be \label{Maxwellbose}
\cL_{eff}\, (\vf) \, =  \, \fr{(-1)^{\, s}}{2\, (s + 1)} \, 
\cR_{\, \m_1 \cdots \m_s, \, \n_1 \cdots \n_s} \, \fr{1}{\Box^{s - 1}} \, \cR^{\, \m_1 \cdots \m_s, \, \n_1 \cdots \n_s} \, ,
\ee 
is nicely expressed as a square of the corresponding curvature, which stresses its role in defining a generalised Maxwell theory.

%%%%%%%%%%%%%%%%%%%%%%%%%%%%%%%%%%%%%%%%%%%%%%%%%%%%%%%%%%%%%%%%%%%%%

\section{Massless fermions}\label{sec:fermions}

%%%%%%%%%%%%%%%%%%%%%%%%%%%%%%%%%%%%%%%%%%%%%%%%%%%%%%%%%%%%%%%%%%%%%

%%%%%%%%%%%%%%%%%%%%%%%%%%%%%%%
\subsection{Irreducible case}\label{sec:irrfermions}
%%%%%%%%%%%%%%%%%%%%%%%%%%%%%%%

 Contrarily to what a naive intuition based on a simple counting of indices might suggest, 
the proper fermionic analogue of the Fierz-Pauli equation \eqref{FP}, capable to serve as a model
for irreducible fermions, is the Rarita-Schwinger equation for a particle of spin $s=3/2$:
\be \label{RS}
\begin{split}
&\dsll \, \psi_{\, \m} \,  - \, \pr_{\, \m} \, \psisl \, = \, 0 \, , \\
& \ \g^{\, \n} \, F_{\, \n \m} (\psi)\, = \,  0 \, \, ,
\end{split}
\ee
where we are omitting the spinor index, and where the second of \eqref{RS}
shows how the same equation obtains taking the gamma-trace of a fermionic field-strength, 
formally identical to Maxwell's tensor:
\be
F_{\, \n \m} (\psi)\, = \, \pr_{\, \n} \, \psi_{\, \m} \, - \,  \pr_{\, \m} \, \psi_{\, \n} \, .
\ee
In Ref.~\citen{fang} Fang and Fronsdal proved that, for a spinor-tensor
$\psi$ of rank $s$, the proper generalisation of the first of \eqref{RS} is
\be \label{S}
\cS \, \equiv \, \dsll \, \psi - \, \pr \, \psisl \, = \, 0 \, ,
\ee
where $\cS$ is gauge invariant under $\d \, \psi = \pr \, \e$ with a gamma-traceless spinor-tensor parameter: 
$\not \! \e \equiv 0$. Under this condition (together with a condition of triple gamma-tracelessness of $\psi$
needed to build a Lagrangian for \eqref{S}) it can be shown that \eqref{S}
propagates  the polarisations pertaining to an irreducible, massless, spin-$(s + 1/2)$ spinor.
However, the constraints prevent the use of the spinorial field-strengths of Ref.~\citen{dwf} and thus
seemingly forbid to generalise the second form of the Rarita-Schwinger equation.
To forego this obstacle one can first  introduce a spinorial compensator field $\xi$ s.t. $\d \xi = \not \!  \! \e$
allowing to define a simple unconstrained generalised Rarita-Schwinger tensor as\cite{fs3, fms1}
\be \label{W}
\cW \, = \, \cS \, + \, 2 \, \pr^{\, 2} \, \xi\, .
\ee
In complete analogy with the bosonic case, the full Lagrangian also involves a Lagrange multiplier 
allowing to formally remove the triple gamma-trace constraint on $\psi$. By manifest gauge invariance, 
the integration over the two auxiliary fields (and over their adjoints) must result in an effective, non-local
Lagrangian depending only on $\psi$ and equivalent to its local antecedent, since no degrees of freedom
are eliminated in this process. In particular for  the case of spin $5/2$ the resulting kinetic tensor looks
\be \label{Wpsi}
\cW_{\, \psi} \, = \, \cS \, + \, 2 \, \pr^{\, 2} \, \xi_{\, \psi} \, = \, \fr{2}{\Box}\, \{\prd {\not \! \! \cR}  \, - \,  \12 \, \fr{\dsll}{\Box}\, \cR^{\, \pe} \, + \,  
\12 \, \fr{\pr^{\, 2}}{\Box^{\, 2}} \, \dsll\, \cR^{\, \pe \pe} \} \, , 
\ee
while the corresponding Lagrangian actually produces the correct propagator.\cite{dariocrete}

Once again, the crucial property of the effective compensator $\xi_{\, \psi} $ defined in \eqref{Wpsi} is not 
just to guarantee gauge invariance of $\cW_{\, \psi}$, 
since to this purpose several other solutions for $\xi_{\, \psi} $ would be available,\cite{dariocrete} 
but instead to ensure that, in the same fashion of its local counterpart \eqref{W},  $\cW_{\, \psi}$ be 
triply-gamma traceless and satisfy the appropriate Bianchi identity.\cite{fms1, dariocrete} 
For spin $s > 5/2$  some informations about the structure of the correct kinetic tensors $\cW_{\, \psi}$ 
can be obtained from the corresponding  mass deformations, as we will  argue in Section \ref{sec:massive}.

%%%%%%%%%%%%%%%%%%%%%%%%%%%%%%%
\subsection{Reducible case}\label{sec:redfermions}
%%%%%%%%%%%%%%%%%%%%%%%%%%%%%%%

Given the role played by the Rarita-Schwinger theory, 
the natural candidate to model
reducible representations is then the Dirac theory for a massless spin $1/2$ particle, defined by the equation 
\be \label{D}
\dsll \, \psi \, = \, 0 \, ,
\ee
where the field $\psi$ itself, by gauge invariance, can be formally considered as being its own curvature.
The proper generalisation of the Dirac theory to higher-spins is then the fermionic triplet\cite{fs2, st}  defined by the equations
\be \label{fermistring}
\begin{split}
& \, \, \cD \, \equiv \,  \dsll \, \psi \, - \, \pr \, \chi \, = \, 0 \, ,  \\
& \dsll \, \chi \, = \, \prd \, \psi \, - \, \pr \, \l \,  , \\
& \dsll \l \, = \, \pr \, \cdot \, \chi \ ,
\end{split}
\ee
linking the unconstrained spinor-tensors $\psi$,  $\chi$ and $\lambda$ in a system invariant under 
$ \d \psi  =  \pr \, \e $, $\d  \chi  = {\not \! \pr}\, \e $ and $\d  \l  = \prd \e $, propagating \emph{all} lower spin 
components in $\psi$ together with the irrep of maximal spin $s + 1/2$. Similarly to its bosonic counterpart,
the system \eqref{fermistring} is also related (with qualifications) to the tensionless limit of the open superstring.
Here we are displaying the full set of equations, in particular to show that both $\chi$ and $\l$ need to be kept at the local level, differently
from the case of the bosonic triplet where the field $C$ had no proper kinetic term. 
However, since neither $\chi$ nor $\l$ carry propagating degrees of freedom
we can still integrate them away without changing the spectrum.  The resulting effective Lagrangian\cite{dariotripl} \be \label{Maxwellfermi}
\cL_{eff}\,(\psi, \pb) \, =  \, \fr{(-1)^{\, s} }{s + 1} \, i\, 
\bar{\cR}_{\, \m_1 \cdots \m_s, \, \n_1 \cdots \n_s} \, \fr{\dsll}{\Box^{s}} \, \cR^{\, \m_1 \cdots \m_s, \, \n_1 \cdots \n_s} \, .
\ee 
displays manifest similarity with its bosonic counterpart \eqref{Maxwellbose}, while the corresponding equations of motion,
\be \label{Dpsi}
\cD_{\, \psi} \, \equiv \,  \dsll \, \psi \, - \, \pr \, \chi_{\, \psi} \, = \, \fr{\dsll}{\Box^{s }} \, \prd^{\, s} \, \cR^{\, (s)} \, (\psi)\,= \, 0\,  ,
\ee
when compared to  \eqref{Tphi}, generalise to any spin the formal relation between the 
Dirac equation \eqref{D} and the Klein-Gordon equation for a massless scalar $\phi$:\cite{fs1}
\be
\dsll \, \psi \, = \, \fr{\dsll}{\Box} \, \Box \, \phi \, ,
\ee
a relation that in such a simple form is never valid for the irreducible case.

%%%%%%%%%%%%%%%%%%%%%%%%%%%%%%%%%%%%%%%%%%%%%%%%%%%%%%%%%%%%%%%%%%%%%

\section{Massive deformation of geometric Lagrangians}\label{sec:massive}

%%%%%%%%%%%%%%%%%%%%%%%%%%%%%%%%%%%%%%%%%%%%%%%%%%%%%%%%%%%%%%%%%%%%%

Consistently with our scheme, we expect higher-spin Lagrangians for massless fields to admit deformations 
at the massive level following a pattern somehow inspired by that of their low-spin models. 
This is only partly realised for their local forms, either constrained or unconstrained.
In both cases it is possible to introduce masses \`a la Stueckelberg by performing a Kaluza-Klein-type
reduction of the massless theory from $d+1$ to $d$ flat space-time dimensions. (With more refined techniques 
one can similarly study massive fields on $d$-dimensional (A)dS spaces embedded in 
$(d+1)$-dimensional Minkowski background.\cite{fms2}) Then, after gauge fixing, 
one can recover other known forms of the massive Lagrangians for higher-spins, as those found by Singh and
Hagen.\cite{SH} The resulting theories always involve auxiliary fields of some sort, 
with cross-couplings proportional to the mass. Differently, for the geometric Lagrangians here reviewed 
the problem of finding proper mass deformations admits much simpler solutions, involving no auxiliary
fields.

%%%%%%%%%%%%%%%%%%%%%%%%%%%%%%%
\subsection{Irreducible case}\label{sec:irrm}
%%%%%%%%%%%%%%%%%%%%%%%%%%%%%%%

Our model in this case is the Fierz-Pauli equation for \emph{massive} spin-$2$ fields:\cite{FP}
\be \label{FPm}
\cR_{\m \n} \, - \, \12 \, \h_{\m \n} \, \cR \, - \, m^{\, 2} \, 
(h_{\m \n} \, - \, \h_{\m \n}h^{\, \a}_{\, \ \a})\, = \, 0 \, ,
\ee
where $\cR_{\m \n}$ and $\cR$ define the linearised Ricci tensor and Ricci scalar, respectively.
The crucial mechanism behind the consistency (and the uniqueness) of the Fierz-Pauli mass term
is the fact that, due to the Bianchi identity satisfied by the Einstein tensor, $\pr^{\, \m}(\cR_{\m \n} -  \12  \h_{\m \n} \cR) \equiv 0$, 
the divergence of \eqref{FPm} leads to the Fierz-Pauli constraint
\be \label{fpc}
\pr^{\, \a} \,  h_{\, \a \, \m} \, - \, \pr_{\, \m} \, h^{\, \a}_{\, \ \a} \, = \, 0 \, ,
\ee
while the latter, in its turn, implies a chain of consequences reducing
\eqref{FPm} to the proper conditions for a massive spin-$2$ particle: $(\Box - m^2) h_{\, \m \, \n} =  0$,  
$\prd h_{\, \m} =  0  =  h^{\, \a}{}_{\, \a}$. While this kind of approach is not suitable for 
theories with constraints or involving auxiliary fields,\ft{In particular because the corresponding
Einstein tensors are not divergenceless.} it does provide the proper
way to proceed for the Lagrangians proposed in Ref.~\citen{fms1}. 
The starting point is the Einstein tensor associated to \eqref{RicciS}, 
\be \label{E}
\cE_{\, \vf} \, = \, \cA_{\, \vf} \, - \, \12 \, \h \, \cA_{\, \vf}^{\, \pe} \, + \, \h^{\, 2} \, \cB_{\, \vf} \, ,
\ee
where the explicit form of $\cB_{\, \vf}$ was given in Ref.~\citen{fms1}, 
together with an ansatz on its massive deformation inspired by \eqref{FPm}:
\be \label{Em}
\cE_{\, \vf} \, \rightarrow \,  \cE_{\, \vf} - \, m^{\, 2} \, M_{\, \vf}  \, ,
\ee
with $M_{\, \vf}$ a linear function of $\vf$ and its traces, to be determined. The solution for $M_{\, \vf}$ is 
then completely fixed asking  that the divergence of the equation of motion lead to the condition
\be \label{gfpc}
\prd \vf - \, \pr \, \vf^{\, \pe} \, = \, 0 \, ,
\ee
which is necessary and sufficient in order for the equation $\cE_{\, \vf} - \, m^{\, 2} \, M_{\, \vf} = 0$ to imply the
Fierz system\cite{fierz} $(\Box - m^2) \vf =  0$,   $\prd \vf =  0  =  \vf^{\, \pe}$.  The generalised Fierz-Pauli mass
term for spin-s bosons so determined is\cite{dario07}
\be \label{FPmb}
M_{\, \vf} \, = \, \vf \, - \, \h \, \vf^{\, \pe} \, - \, \sum_{k = 2}^{[\fr{s}{2}]} \, \fr{1}{(2 k - 3)! !} \, \h^{\, k} \, \vf^{\, [k]} \, ,
\ee
where one can appreciate that the standard Fierz-pauli mass term, $\vf \, - \, \h \, \vf^{\, \pe}$, is just the beginning
of a sequence involving all possible traces of $\vf$. Let us also observe that the coefficients in $M_{\, \vf}$
depend neither on the spin nor on the space-time dimension.

Analogous arguments also apply to fermions. In this case the model is the massive Rarita-Schwinger equation
\be
\cS \, - \, \12 \, \g \, \ssl \, + \, 
m \, (\psi \, - \, \g \, \psisl) \, = \, 0\, ,
\ee
leading to the on-shell condition $\prd \psi \, - \, {\not \! \pr} \,  {\not \! \!  \psi} \, = \, 0$.
However, in Ref.~\citen{dario07} it was found that, for higher-rank spinor-tensors, assuming that  fermionic Einstein tensors
exist satisfying $\prd \cE_{\, \psi} \, \equiv \, 0$ (as those constructed in Ref.~\citen{fs1}), then the 
proper requirement to be imposed is that $\prd M_{\, \psi} = 0$ imply the more general fermionic-Fierz-Pauli constraint
\be \label{fpcf}
\prd \psi \, - \, \dsll \, \psisl \, - \, \pr \, \psi^{\, \pe} \, = \, 0 \, .
\ee
Imposing that \eqref{fpcf} be recovered on-shell one finds for $M_{\, \psi}$ the unique solution 
\be \label{FPmf}
M_{\, \psi} \, = \, \psi \, - \, \g \, \psisl \, - \, \h \, \psi^{\, \pe} \, - \,  \sum_{k = 1}^{[\fr{s -1}{2}]} \, \fr{1}{(2 k - 1)! !}
\, \g \, \h^{\, k} \, \psisl^{\, [k]} \, - \, 
\sum_{k = 2}^{[\fr{s}{2}]} \, \fr{1}{(2 k - 3)! !}
\, \h^{\, k} \, \psi^{\, [k]} \, .
\ee
A couple of additional comments are in order:
\begin{itemize}
\item As observed in Refs.~\citen{dario07, darioValencia}, the form of the propagator is \emph{completely determined by 
the coefficients of the mass term}. This is already true for the spin-$2$ case and makes it possible to find $M_{\, \vf}$ 
($M_{\, \psi}$) only exploiting the fact that $\cE_{\, \vf}$ ($\cE_{\, \psi}$)  be identically divergenceless, 
regardless of whether or not the kinetic term  itself defines the correct propagator for $m = 0$.
\item However, as anticipated, the structure of $M_{\, \vf}$ ($M_{\, \psi}$) do carry information on the proper massless theory since 
its coefficients (consistently with the Kaluza-Klein mechanism) dictate the form of the term in $\Box$
(or $\not \! \! \pr$) for the massless theory associated to  the correct propagator. 
\end{itemize}

%%%%%%%%%%%%%%%%%%%%%%%%%%%%%%%
\subsection{Reducible case}\label{sec:redm}
%%%%%%%%%%%%%%%%%%%%%%%%%%%%%%%

Here we would like to discuss the mass deformation of Lagrangians \eqref{Maxwellbose} and \eqref{Maxwellfermi};
this extension was not previously considered. Once again, our aim is to  get inspiration from the massive 
deformations of the corresponding low-spin models.
While this is partly achieved for the irreducible case, where anyway more structures need to be introduced as the
spin increases, it is enticing to observe that the correspondance in the reducible case applies almost verbatim.

In fact, looking at the Proca Lagrangian for a massive vector
\be \label{PL}
\cL  \, = \, - \, \fr{1}{4} \, F_{\, \m \n}\, F^{\, \m \n} \, - \, \12 \, m^{\, 2} \, A_{\, \m} \,  A^{\, \m} \, ,
\ee
we are led to conjecture the following \emph{generalised Proca Lagrangian} 
\be \label{Pphi}
\cL \, = \, \, \fr{(-1)^{\, s}}{2\, (s + 1)} \, 
\cR_{\, \m_1 \cdots \m_s, \, \n_1 \cdots \n_s} \, \fr{1}{\Box^{s - 1}} \, \cR^{\, \m_1 \cdots \m_s, \, \n_1 \cdots \n_s}
 \, \, - \, \12 \, m^{\, 2} \, \vf^{\, 2} \, .
\ee
whose consistency is indeed easily checked, since the divergence of the corresponding equation
\be  \label{eqPphi}
\fr{1}{\Box^{s - 1}} \, \prd^{\, s} \, \cR \, -  \, m^{\, 2} \, \vf  \, = \, 0 \, ,
\ee
can be shown to imply\ft{There is a simple, group theoretical reason for that: computing $s+1$ divergences 
of $\cR$ forces the symmetrization of one index in a group with all the indices of the other group; the resulting tensor then 
vanishes identically because of irreducibility of the curvature as a GL(D) tensor.}  $\prd \vf = 0$, while under this latter
condition the whole tensor $\fr{1}{\Box^{s - 1}} \, \prd^{\, s} \, \cR$ collapses to $\Box \, \vf$, finally providing the appropriate
reduced Fierz system $(\Box \, - \, m^{\, 2})\, \vf=0$ and $\prd \vf = 0$.

 Similarly for fermions the model of the massive Dirac Lagrangian
\be \label{DL}
\cL \,=\,   i \, \bar{\psi}\, (\dsll \, - \, m)\, \psi \, ,
\ee 
gets generalised to
\be \label{Dpsi}
\cL \, =  \, \fr{(-1)^{\, s} }{s + 1} \, i\, 
\bar{\cR}_{\, \m_s, \,  \n_s} \, \fr{\dsll}{\Box^{s}} \, \cR^{\,  \m_s, \,  \n_s}  \, - \, i \, m \, \pb \, \psi \, ,
\ee
while the same argument given for bosons ensures the formal consistency of the resulting equations of motion.

Our derivation of the mass deformations for the various geometric Lagrangians
did not arise from eliminating auxiliary fields in the corresponding local theories. 
If we were to follow that procedure the corresponding massive Lagrangians
would probably look more complicated than those here described. However, because of the consistency
of our massive theories on-shell, and in view of the correctness of their current exchanges, we believe that
the possible corrections to \eqref{Pphi} and \eqref{Dpsi} should reduce to terms on-shell irrelevant, like
gauge artifacts associated with the Stueckelberg symmetry emerging after Kaluza-Klein reduction.

%%%%%%%%%%%%%%%%%%%%%%%%%%%%%%%%%%%%%%%%%%%%%%%%%%%%%%%%%%%%%%%%%%%%%

\section{Outlook}\label{sec:conclusions}

%%%%%%%%%%%%%%%%%%%%%%%%%%%%%%%%%%%%%%%%%%%%%%%%%%%%%%%%%%%%%%%%%%%%%

A number of gaps are still to be filled in the scheme we proposed in this contribution. To begin with, for symmetric bosons
(and fermions), and more generally for  (spinor-)tensors of mixed symmetry in flat background, there is at present no counterpart 
of the constrained (Fang-)Fronsdal formulation describing the \emph{reducible} particle content of the triplets \eqref{triplet}, \eqref{fermistring}. 
In addition, for the mixed-symmetry case, the constrained equations 
of Labastida\cite{labastida} were recently given a  Lagrangian derivation
for fermions and then generalised to their minimal unconstrained  version\cite{cfms}, 
but the geometric theory that would result after integrating over the  auxililary fields is still to be computed.
Much less is known for (A)dS spaces, for which a full Lagrangian description 
in metric form is still missing even for bosons. Some of these issues will be discussed in a forthcoming paper.\cite{AD}

\section*{Acknowledgements}
I am grateful to A. Campoleoni, J. Mourad and A. Sagnotti for collaboration on several
topics discussed in this review, and to X. Bekaert, T. Erler and M. Schnabl for stimulating
discussions. I would like to thank the Yukawa Institute for Theoretical Physics at
Kyoto University, for the opportunity to present my work at the 
YITP-W-10-13 on "String Field Theory and Related Aspects".
This work was supported  by the Grant-in-Aid for the Global COE Program
"The Next Generation of Physics, Spun from Universality and
Emergence" from the Ministry of Education, Culture, Sports, Science and
Technology (MEXT) of Japan, and
by the EURYI grant EYI/07/E010 from EUROHORC and ESF.

\end{document}